\documentclass[aps,10pt,final,
notitlepage, oneside, twocolumn, nobibnotes, nofootinbib,
superscriptaddress,showpacs, centertags, showkeys,  assume]{revtex4}

\usepackage{graphicx}
\usepackage{amsmath}

\begin{document}

\title{Phenomenological insight into JLab proton polarization data puzzle by deuteron
impulse approximation}
\date{\today}

\author{C.Adamu\v s\v c\'in}
\address{Institute of Physics, Slovak Academy of Sciences,
Bratislava, Slovak Republic}
\author{L.Bimbot}
\address{IPNO, IN2P3, BP 1, 91406 Orsay, France}
\author{S.Dubni\v cka}
\address{Institute of Physics, Slovak Academy of Sciences,
Bratislava, Slovak Republic}
\author{A.Z.Dubni\v ckov\'a}
\address{Dept. of Theoretical Physics, Comenius Univ., Bratislava,
Slovak Republic}
\author{E.Tomasi-Gustafsson}
\address{DAPNIA/SPhN, CEA/Saclay, 91191 Gif-sur-Yvette Cedex,
France}

\vspace{2cm}
\begin{abstract}
The non-relativistic impulse approximation of deuteron
electromagnetic form factors is used to investigate the space-like
region behavior of the proton electric form factor in regard of
the two contradictory experimental results extracted either from
Rosenbluth separation method or from recoil proton JLab
polarization data.
\end{abstract}

\pacs{12.20.-m,13.40.-f, 13.60.Hb, 13.88.+e}
\keywords{electromagnetic interactions, polarization,  electromagnetic
formfactors, deuteron}
\maketitle

\section{Introduction}
\hspace{0.7cm}

A study of the deuteron electromagnetic (EM) structure is very
interesting as the deuteron is the most simple bound state system
of nucleons, which provides an excellent opportunity to study
nucleon-nucleon (NN) interaction as well as its dependence on
electromagnetic structure of underlying nucleons. Up to now
several (phenomenological) models and fits of the deuteron
electromagnetic structure have been developed. The purely
phenomenological fits \cite{Ab00} give small $\chi^2$ but they
come without any physical background. The vector meson dominance
(VMD) models (and their generalization by incorporating correct
deuteron form factor (FF) analytic properties)
\cite{Du91,Ko95,To06} describe deuteron EM structure through
exchange of isoscalar vector mesons $\omega, \phi$ (and their
excitations) and they don't assume NN interaction effects
explicitly. Another class of models, non-relativistic (NIA) and
relativistic (RIA) impulse approximations, reviewed in
\cite{Gi02}, assume NN interaction based on deuteron wave
functions and describe the deuteron EM FFs through the nucleon EM
FFs. Just such models seem to be able to bring some light into
existence of two contradicting space-like behaviors of the
electric nucleon FF, obtained in elastic scattering of unpolarized
electrons on unpolarized protons by Rosenbluth method
and in polarization transfer process $\vec{e}%
^{-}p\rightarrow e^{-}\vec{p}$  measuring transversal and
longitudinal components of the recoil proton's polarization
simultaneously.

\section{Two contradicting proton electric form factor behaviors in space-like region}
\hspace{0.7cm}

The electron-proton elastic scattering was used to be the most
common way of the proton electromagnetic structure study from the
half of 50's of the last century \cite{Hof58} and abundant data
(more than 400 data points) on proton electric $G_{Ep}(t)$ and
magnetic $G_{Mp}(t)$ FFs in the space-like region $-Q^2=q^2=t<0$
appeared (for references see paper \cite{Du03}). They have been
obtained from the measured differential cross-section of the
elastic scattering of unpolarized electrons on unpolarized protons
in laboratory system
\begin{eqnarray}
\nonumber \frac{d\sigma^{lab}(e^-p\to
e^-p)}{d\Omega}=\frac{\alpha^2}{4E^2}\frac{\cos^2(\theta/2)}{\sin^4(\theta/2)}
\frac{1}{1+(\frac{2E}{m_p})\sin^2(\theta/2)}\nonumber
\end{eqnarray}
\begin{equation}
.\left[A(t)+B(t)\tan^2(\theta/2)\right], \label{difcs}
\end{equation}

$\alpha=1/137$, $E$-the incident electron energy\\
\begin{eqnarray}
 A(t)=\frac{G^2_{Ep}(t)-\frac{t}{4m_p^2}G^2_{Mp}(t)}{1-\frac{t}{4m_p^2}},
 \quad
 B(t)=-2\frac{t}{4m_p^2}G^2_{Mp}\label{a2}
\end{eqnarray}
by the Rosenbluth technique. Their ratio
$\mu_{p}G_{Ep}(t)/G_{Mp}(t)$ in error bars roughly equals one,
showing the electric and magnetic distributions in the proton to
be equal.

   Recently in JLab \cite{Jon00}-\cite{Pun05}, measuring transverse
\begin{eqnarray}
P_t=\frac{h}{I_0}(-2)\sqrt{\tau(1+\tau)}G_{Mp}G_{Ep}\tan(\theta/2)\label{a3}
\end{eqnarray}
and longitudinal
\begin{eqnarray}
P_l=\frac{h(E+E')}{I_0m_p}\sqrt{\tau(1+\tau)}G^2_{Mp}\tan^2(\theta/2)\label{a4}
\end{eqnarray}
components of the recoil proton's polarization (as suggested in
Refs. \cite{Akh68}) in the electron scattering plane of the
polarization transfer process $\vec{e} ^{-}p\rightarrow
e^{-}\vec{p}$ ($h$ is the electron beam helicity, $I_{0}$ is the
unpolarized cross-section excluding $\sigma _{Mott}$ and $\tau
=-t/4m_{p}^{2}$) simultaneously, the very precise and surprising
data on the ratio $\mu_{p}G_{Ep}(t)/G_{Mp}(t)$ have been obtained,
showing the electric and magnetic distributions in the proton to
be different, contrary to what was followed from Rosenbluth data.

   The latter contradiction is now well known as the JLab proton
polarization data puzzle and a natural question is arisen "Which
data are correct"?

   A lot of effort has been devoted to the search of a definite solution
\cite{Afa01}-\cite{Bys06}. The explanation is likely to be related
to radiative corrections, which may reach $40\%$ on the
unpolarized cross-section, but are negligible on the polarization
ratio.

   In Ref. \cite{Afa01} the structure functions method was applied
for an evaluation of the radiative corrections in transferred
polarization experiment and their size was of order of magnitude
too small to bring the polarization data in agreement with the
Rosenbluth ones.

   In the papers \cite{Gui03}, \cite{Blu05} authors came
to a conclusions that the presence of the two photon contributions
could solve the problem, however, the kinematical properties
related to fast decreasing form factors and the possible presence
of inelastic contributions in the intermediate state were not
investigated in detail.

   The structure functions method has been also applied in Ref.
\cite{Du05}, where it was shown that the corrections can become
very large, if one takes into account the initial state photon
emission. However, the corresponding kinematical region is usually
rejected (by appropriate selection of the scattered electron
energy) in the experimental analysis.

   In the paper \cite{Afa05} a model dependent calculation based
on GPD showed an agreement more qualitative than quantitative with
one set of data \cite{Andi94}.

   In Ref. \cite{To05} it was noted that the reason of the
discrepancy lies in the slope of the reduced cross-section as a
function of the virtual photon polarization.

   In the paper \cite{Bys06} higher order radiative corrections for
initial state emission, in frame of the structure function
approach, has been applied to polarized and unpolarized
cross-section. It was shown that they have the right size and
behaviour to bring data in agreement. The $2\gamma$ photon box is
shown to be irrelevant to this problem in the approximation of
rapidly decreasing FFs and an equally shared momentum between the
two photons.

   In this work we take another approach, a phenomenological point of view
to investigate this problem. We show that the proton electric FF
$G_{Ep}(t)$, and in no case the magnetic FF, is responsible for
the discrepancy in the above-mentioned sets of results.

   For the latter there are indications already in the form of the differential
cross-section (\ref{difcs}), where the proton magnetic FF is
multiplied by $-t/(4m^2_p)$ factor, i.e. as $-t$ increases, the
measured cross-section (1)  becomes dominant by $G^2_{Mp}(t)$ part
contribution, making the extraction of $G^2_{Ep}(t)$ more and more
difficult. As a result,  one can have confidence only in the
proton magnetic FF data obtained by the Rosenbluth technique in
higher values of momentum transfer squared.

   A definite conclusion for the above-mentioned assertion follows
from the global analysis \cite{Du04}, \cite{Ad05} of the nucleon
FF data in the framework of the ten-resonance unitary and analytic
model of the nucleon EM structure \cite{Du03}, which provides a
test of:
\begin{itemize}
\item
consistency of the JLab proton polarization data with all other
proton and neutron EM FF data
\item
consistency of these data with the powerful tool of physics - the
analyticity
\item
their consistency with the asymptotic behaviour as predicted (up
to logarithmic corrections) by QCD.
\end{itemize}

\begin{figure*}
    \centering
    \scalebox{0.7}{\includegraphics{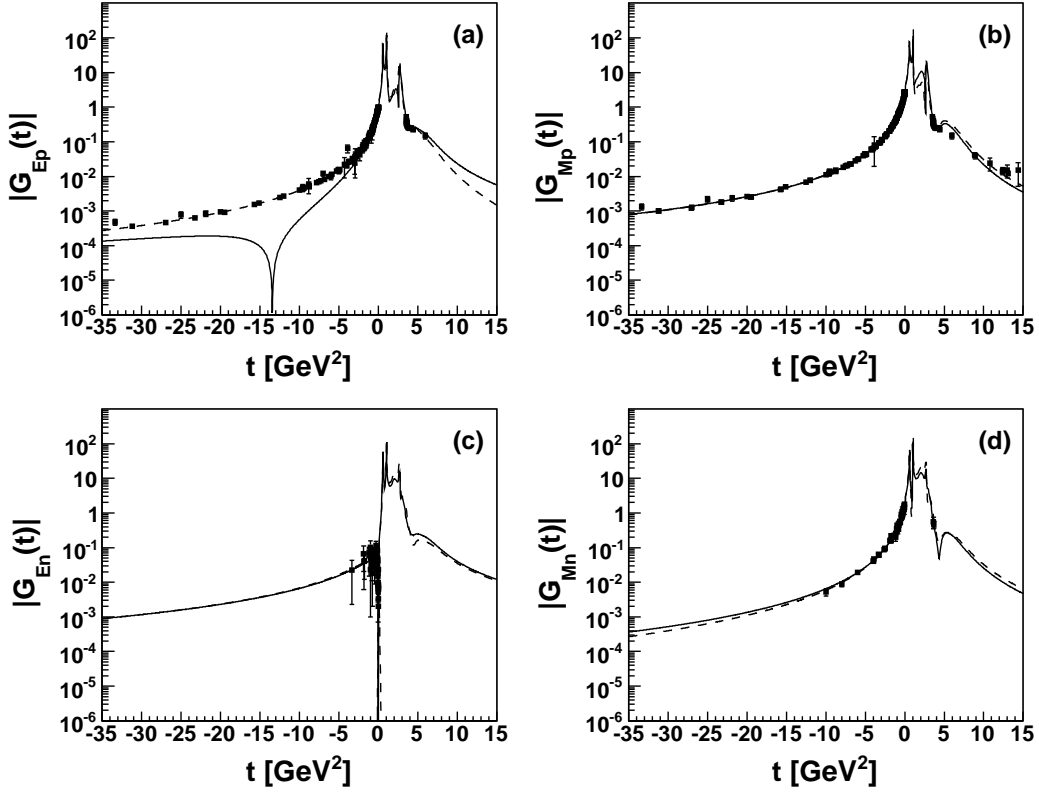}}
    \caption{\small{Results of the analysis of Rosenbluth (dashed lines)
    and JLab (full lines) data by nucleon EM structure U\&A model.}}
    \label{fig:nucFFs}
\end{figure*}

   In the analysis all Rosenbluth data for proton electric FF in space-like
region were substituted for the ratio $\mu_{p}G_{Ep}(t)/G_{Mp}(t)$
data obtained in JLab proton polarization experiments. As a
result, the parameters in comparison with those in \cite{Du03}
were changed very little, a perfect description of the new JLab
data was achieved and almost nothing has been changed in the
description of $G_{Mp}(t)$, $G_{En}(t)$, $G_{Mn}(t)$ in both the
space-like and time-like regions and $|G_{Ep}(t)|$ in the
time-like region. However, the new JLab data strongly require an
existence of the zero (see the full line in Fig.1a), e.i. a
diffraction minimum in the space-like region of $|G_{Ep}(t)|$
around $t=-Q^2=-13~ GeV^2$.

   Consequently, there are two contradicting behaviors (see Fig.1a)
of the proton electric FF $G_{Ep}(t)$ in the space-like region. In
the following sections they are tested in the framework of the NIA
by existing deuteron electromagnetic structure data.

\section{Non-relativistic impulse approximation for deuteron EM structure}

\hspace{0.7cm}

  In the one-photon exchange approximation the deuteron EM structure
can be described by three scalar functions - EM FFs \cite{Ca61}
and the matrix element of the deuteron EM current can be written,
in the most general form, as
\begin{eqnarray}
\nonumber&-&\left\langle d'\left\vert J_{\mu }^{EM}\right\vert
d\right\rangle = G_{1}(q^{2})(\xi ^{\prime \ast }.\xi )d_{\mu
}+G_{2}(q^{2}) \left[ \xi _{\mu }(\xi ^{\prime \ast }.q)-\right.\\
&-&\left.\xi _{\mu }^{\prime \ast }(\xi .q) \right]
-G_{3}(q^{2}).\frac{(\xi .q)(\xi ^{\prime \ast }.q)}{2m_{d}^{2}}
d_{\mu },\label{a5}
\end{eqnarray}
where $\xi ,\xi ^{\prime }$ are polarization vectors of incoming
and outgoing deuterons $d,d'$ of four-momenta $p_{\mu },p_{\mu
}^{\prime }$ obeying the relations
\begin{eqnarray*}
&&\xi ^{\prime }.p^{\prime }=0~;~\xi .p=0~;~\xi ^{\prime 2}=-1~;
~\xi^{2}=-1~;~d_{\mu }=p_{\mu }^{\prime }+p_{\mu }~;\\
 &&~q_{\mu }=p_{\mu}^{\prime }-p_{\mu }. \nonumber
\end{eqnarray*}
However, from practical point of view another linear combinations
of deuteron EM FFs are used \cite{Gi02}
\begin{eqnarray}
G_{C}(q^{2}) &=&G_{1}(q^{2})+\frac{2}{3}\eta\label{a6}
G_{Q}(q^{2})\nonumber \\
G_{M}(q^{2}) &=&G_{2}(q^{2})\\
\label{deuffs0} G_{Q}(q^{2}) &=&G_{1}(q^{2})-G_{2}(q^{2})+(1+\eta
)G_{3}(q^{2})  \nonumber
\end{eqnarray}
where $\eta =\frac{-q^{2}}{4m_{d}^{2}}$, $G_{C}(q^{2})$ is
deuteron charge form factor, $G_{M}(q^{2})$ is deuteron magnetic
form factor and $G_{Q}(q^{2})$ is deuteron quadrupole form factor.

The calculation of deuteron EM FFs within impulse approximation
requires a knowledge of the deuteron wave function and nucleon EM
FFs. As deuteron can be found in S- ($\approx 96\%$) and D-state
($\approx 4\%$), then NN non-relativistic full wave function of
the deuteron can be written in terms of two scalar wave functions

\begin{eqnarray}
\Psi _{abm}
&=&\sum_{l}\sum_{m_{s}}\frac{z_{l}(r)}{r}Y_{l,m-m_{s}}(\widehat{
\mathbf{r}})\chi _{ab}^{1m_{s}}\nonumber \\
&&\left\langle l,1,m-m_{s},m_{s}|1,m\right\rangle=  \nonumber \\
&=&\frac{u(r)}{r}Y_{0,0}(\widehat{\mathbf{r}})\chi
_{ab}^{1m}\label{a7}
\\ &+&\frac{w(r)}{r} \sum_{m_{s}}Y_{2,m-m_{s}}\chi
_{ab}^{1m_{s}}\left\langle 2,1,m-m_{s},m_{s}|1,m\right\rangle
,\nonumber
\end{eqnarray}
where $\left\langle l,1,m-m_{s},m_{s}|1,m\right\rangle $ are
Clebsh-Gordan coefficients, $Y_{l,m_{l}}$ are spherical harmonics
normalized to unity on the unit sphere and $z_{0}=u$, $z_{2}=w$
are reduced $S-$ and $D$-state wave functions, respectively.

The normalization condition
\begin{equation}
\int d^{3}r\Psi _{abm^{\prime }}^{\dag }\Psi _{abm}=\delta _{m^{\prime }m} \nonumber
\end{equation}
implies normalization
\begin{equation}
\int_{0}^{\infty }dr\left[ u^{2}(r)+w^{2}(r)\right] =1,\label{a8}
\end{equation}
which could be understood as probability of finding deuteron in
$S-$ or $D$-state. The $D$-state probability
\begin{equation}
P_{D}=\int_{0}^{\infty }drw^{2}(r) \nonumber
\end{equation}
is an interesting measurement of the strength of the tensor component of the $NN$
force.

\begin{figure}
    \centering
\scalebox{0.85}{\includegraphics{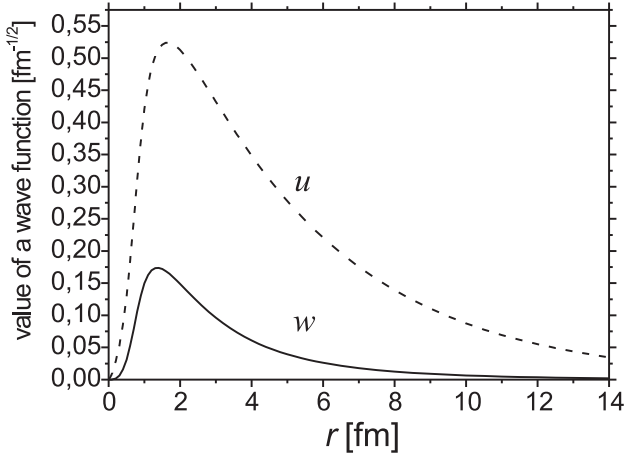}}
    \caption{\small{The $S-$ and $D-$state wave functions ($u$ and $w$, respectively)
    behaviors for Paris potential.}}
    \label{fig:wavefunctions}
\end{figure}

The best non-relativistic wave functions are calculated from the
Schr\"{o}dinger equation using a potential adjusted to fit NN
scattering data for lab energies from 0 to 350 MeV. In this paper
we will use one of the most common potentials called \textit{Paris
potential} \cite{La80}, which was among the first potentials to be
determined from such realistic fit. The $S-$ and $D-$state wave
functions determined from this model are presented in Fig.
\ref{fig:wavefunctions}.

The deuteron is an isoscalar target, therefore within
non-relativistic impulse approximation, its FFs depend only on the
isoscalar nucleon form factors $G_{EN}^s$ and $G_{MN}^s$
\begin{eqnarray}
G_{EN}^s&=&G_{Ep}+G_{En} \nonumber \\
G_{MN}^s&=&G_{Mp}+G_{Mn} \label{nuciso}
\end{eqnarray}
in the following way
\begin{eqnarray}
G_{C} &=&G_{EN}^{s}D_{C}  \nonumber \\
G_{M} &=&\frac{m_d}{2m_p}\left[
G_{MN}^{s}D_{M}+G_{EN}^{s}D_{E}\right]\label{deutffs}\\
G_{Q}
&=&G_{EN}^{s}D_{Q}  \nonumber
\end{eqnarray}
where the body form factors $D_{C}$, $D_{M}$, $D_{E}$ and $D_{Q}$
are functions of the momentum transfer squared $t$. The
non-relativistic formulas for the body form factors $D$ involve
overlaps of the wave functions $u(r),w(r)$, weighted by spherical
Bessel functions
\begin{eqnarray}
D_{C}(q^{2}) &=&\int_{0}^{\infty }dr\left[
u^{2}(r)+w^{2}(r)\right]
j_{0}(\kappa )  \nonumber \\
D_{M}(q^{2}) &=&\int_{0}^{\infty }dr\left[ 2u^{2}(r)-w^{2}(r)\right]
j_{0}(\kappa)\nonumber \\
&+&\left[ \sqrt{2}u(r)w(r)+w^{2}(r)\right] j_{2}(\kappa)  \nonumber \\
D_{E}(q^{2}) &=&\frac{3}{2}\int_{0}^{\infty }drw^{2}(r)\left[
j_{0}(\kappa)+j_{2}(\kappa)\right]\\ \label{bodyffs}
D_{Q}(q^{2})
&=&\frac{3}{\sqrt{2}\eta }\int_{0}^{\infty }dr w(r)\left[ u(r)-
\frac{w(r)}{\sqrt{8}}\right] j_{2}(\kappa )  \nonumber
\end{eqnarray}
where $\kappa =qr/2$. At $q^{2}=0,$ the body form factors become
\begin{eqnarray}
D_{C}(0) &=&\int_{0}^{\infty }dr\left[ u^{2}(r)+w^{2}(r)\right] =1  \nonumber \\
D_{M}(0) &=&\int_{0}^{\infty }dr\left[ 2u^{2}(r)-w^{2}(r)\right]
=2-3P_{D}
\nonumber \\
D_{E}(0) &=&\frac{3}{2}\int_{0}^{\infty
}drw^{2}(r)=\frac{3}{2}P_{D}\\
D_{Q}(0) &=&\frac{m_{d}^{2}}{\sqrt{50}}\int_{0}^{\infty }drw(r)\left[ u(r)-%
\frac{w(r)}{\sqrt{8}}\right]  \nonumber
\end{eqnarray}
giving the non-relativistic predictions
\begin{eqnarray}
Q_{d} &=&D_{Q}(0)  \label{a9} \\
\mu _{d} &=&\mu _{N}^s D_{M}(0)+D_{E}(0)=\mu
_{N}^s(2-3P_{D})+1.5P_{D}\nonumber,
\end{eqnarray}%
where $Q_{d}$ is the quadrupole moment of the deuteron, $\mu _{d}$
is the magnetic moment of the deuteron and $\mu
_{N}^s=\frac{1}{2}(\mu_p+\mu_n-1)$ is the isoscalar nucleon
magnetic moment. The experimental value of the deuteron magnetic
moment $\mu _{d}=1.7139$, leads to probability of $D$-state
$P_{D}=4.0\%.$ But this is only approximate value, because the
magnetic moment is very sensitive to relativistic corrections.

   Experimentally the EM structure of the deuteron is
measured in the elastic scattering of electrons on deuterons,
described by the differential cross-section (\ref{difcs}) with

\begin{eqnarray}
 A(t)&=& G^2_{C}(t)+\frac{2}{3}\eta G^2_{M}(t)+\frac{8}{9}\eta^2
 G^2_{Q}(t),\nonumber \\
 B(t)&=&\frac{4}{3}\eta(1+\eta)^2G^2_{M}(t),\label{a10}
\end{eqnarray}
and applying the Rosenbluth technique. As a result the data on
structure functions $A(t)$, $B(t)$ are obtained, which are found
to be compiled in the paper \cite{Gi02}.

\section{Results}

\begin{figure}
    \centering
        \scalebox{0.4}{\includegraphics{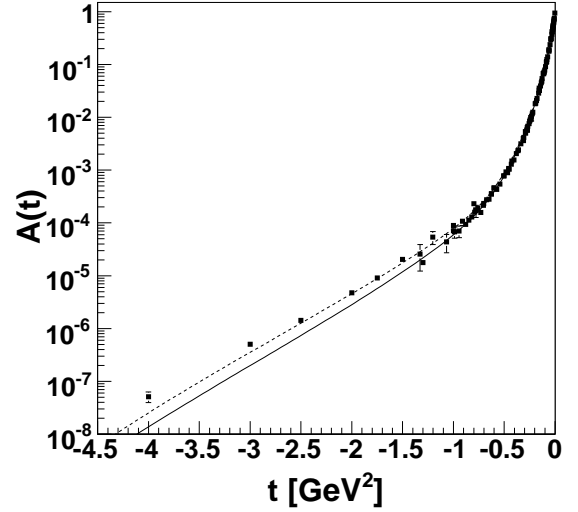}}
    \caption{\small{Deuteron structure function $A(t)$.}}
    \label{fig:A}
\end{figure}

\begin{figure}
    \centering
        \scalebox{0.4}{\includegraphics{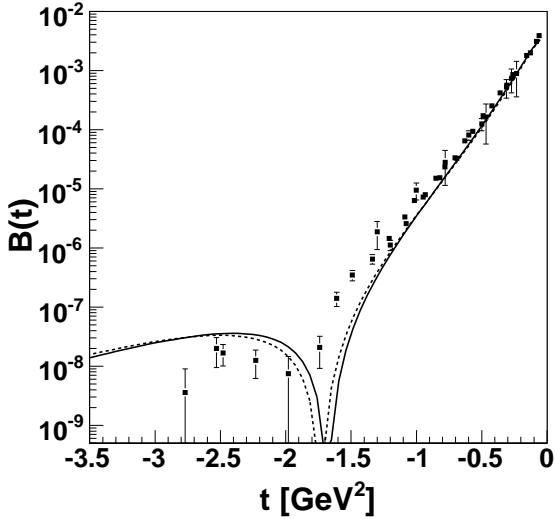}}
    \caption{\small{Deuteron structure function $B(t)$.}}
    \label{fig:B}
\end{figure}

   In order to test the two contradicting behaviors of
$G_{Ep}(t)$, as shown in Fig.1a, in comparison with the deuteron
structure functions $A(t)$, $B(t)$ data, we use expressions for
deuteron EM FFs (\ref{deutffs}) to be expressed through nucleon EM
FFs. First, the fits of Rosenbluth data were made within Unitary
and Analytic model of nucleon EM FFs \cite{Du03}, then the fits of
JLab proton polarization data with all other existing nucleon EM
FFs data. From both behaviors the isoscalar nucleon FFs were
determined, by means of which the deuteron FFs have been found and
as a result the two differen behaviors of deuteron structure
functions $A(t)$, $B(t)$ were calculated.

\begin{figure}
    \centering
        \scalebox{0.4}{\includegraphics{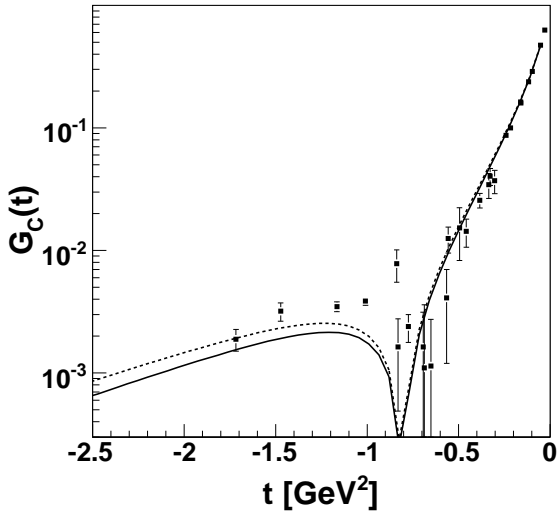}}
    \caption{\small{Charge deuteron FF $G_C(t)$.}}
    \label{fig:Gc}
\end{figure}

\begin{figure}
    \centering
        \scalebox{0.4}{\includegraphics{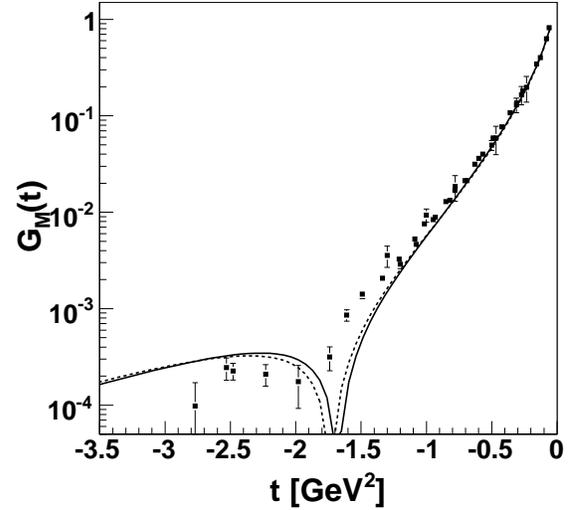}}
    \caption{\small{Magnetic deuteron FF $G_M(t)$.}}
    \label{fig:Gm}
\end{figure}

\begin{figure}
    \centering
        \scalebox{0.4}{\includegraphics{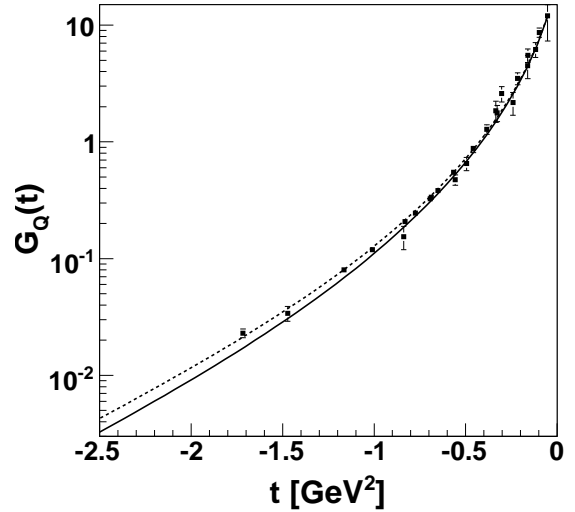}}
    \caption{\small{Quadrupole deuteron FF $G_Q(t)$.}}
    \label{fig:Gq}
\end{figure}
By a comparison of the latter (see Figs. 3 and 4) with existing
deuteron structure functions data corresponding $\chi^2s$ are
calculated as presented in Table \ref{tablechi},

\begin{table}[htbp]
    \centering
        \begin{tabular}{|c|c|c|}
        \hline
        &$\chi^2_{A}$&$\chi^2_{B}$\\
        \hline
        JLab&926&476\\
        \hline
        Rosenbluth&2080&573\\
        \hline
        \end{tabular}
    \caption{The $\chi^2$ of deuteron structure functions $A(t),B(t)$ for two
different scenarios.} \label{tablechi}
\end{table}

from where one can see immediately that the behaviors of $A(t)$,
$B(t)$, obtained by means of the $G_{Ep}(t)$ with the zero around
$t=-13~ GeV^2$ unambiguously is preferred.

   For completeness we present (see Figs. 5, 6 and 7) also the obtained behaviors of
deuteron EM FFs $G_{C}(t)$, $G_{M}(t)$ and $G_{Q}(t)$ with the
data \cite{Gi02} obtained in recent polarization experiments.

\section{Conclusions}

  The Rosenbluth data on the proton electric FF in
space-like region and the new JLab proton polarization data on the
ratio $\mu_{p}G_{Ep}(t)/G_{Mp}(t)$, were both analyzed together
with all other existing nucleon FFs data in space-like and
time-like regions by means of the ten-resonance unitary and
analytic nucleon EM structure model. As a result, two
contradicting behaviors of $G_{Ep}(t)$ in space-like region were
obtained. They are brought into a comparison with other
independent data, the data on deuteron structure functions, by
means of the non-relativistic impulse approximation of deuteron EM
structure. From the values of $\chi^2$ shown in Table I it
unambiguously follows that the $G_{Ep}(t)$ from the JLab proton
polarization data analysis with the zero around $t=-13~ GeV^2$ are
more consistent with the deuteron structure functions $A(t),B(t)$
data than the older Rosenbluth behavior.

The work was partly supported by Slovak Grant Agency for Sciences
VEGA, Grant No. 2/4099/26 (C.A., S.D. and A.Z.D.).

\end{document}